\documentclass[aps,prd,floats,floatfix,twocolumn,nofootinbib,showpacs]{revtex4}
\usepackage{amsmath} 
\usepackage{amssymb} 
\usepackage{graphicx}
\usepackage{bm}

\newcommand{\gfour}{{^{(4)}{\mathbf g}}}

\newcommand{\CF}{\psi}
\newcommand{\trK}{K}

\newcommand{\g}{g}
\newcommand{\cg}{\tilde{g}}
\newcommand{\K}{K}

\newcommand{\cA}{\tilde{A}}

\newcommand{\cderiv}{\tilde{\nabla}}

\newcommand{\cLong}{\tilde{\mathbb{L}}}
\newcommand{\R}{R}
\newcommand{\cR}{\tilde{R}}
\newcommand{\N}{N}
\newcommand{\cN}{\tilde{N}}

\newcommand{\cu}{\tilde{u}}

\begin{document}

\title{Initial data for Einstein's equations with superposed
gravitational waves}

\author{Harald P. Pfeiffer${}^1$, Lawrence E. Kidder${}^2$, Mark
A. Scheel${}^1$, and Deirdre Shoemaker${}^3$ }

\affiliation{${}^1$ Theoretical Astrophysics 130-33, California
Institute of Technology, Pasadena, CA 91125} \affiliation{${}^2$
Center for Radiophysics and Space Research, Cornell University,
Ithaca, NY 14853} \affiliation{${}^3$ Department of Physics and
Institute of Gravitational Physics and Geometry, Penn State
University, University Park, PA 16802}

\date{\today}

\begin{abstract}
A method is presented to construct initial data for Einstein's
equations as a superposition of a gravitational wave perturbation on
an arbitrary stationary background spacetime.  The method combines the
conformal thin sandwich formalism with linear gravitational waves, and
allows detailed control over characteristics of the superposed
gravitational wave like shape, location and propagation direction.
It is furthermore fully covariant with respect to spatial coordinate
changes and allows for very large amplitude of the gravitational wave.
\end{abstract}


\maketitle

\section{Introduction}

Vacuum spacetimes which are perturbed away from stationary solutions
of Einstein's equations are interesting in many different aspects.
Using perturbations of flat space, one can examine (critical) collapse
to black holes~\cite{Choptuik:1993,Abrahams-Evans:1993}, or
investigate nonlinear interaction between gravitational waves.
Perturbed black holes are expected to be produced by astrophysical
events like binary black hole coalescence.  Detailed understanding of
the behavior of perturbed black holes, including the nonlinear regime,
will be important for analyzing data from gravitational wave detectors like
GEO, LIGO, TAMA and VIRGO.  Numerical evolutions are the only known
avenue to analyze Einstein's equation in general, three-dimensional,
nonlinear situations.  Moreover, non-stationary spacetimes without
black holes or with just one perturbed black hole provide important
testbeds and benchmarks for numerical evolution codes in a
computational setting much simpler than a full binary black hole
evolution.

Such numerical evolutions require initial data representing perturbed
spacetimes.  Historically, Brill waves~\cite{Brill:1959} are the most
widely used approach to construct perturbations of Minkowski space
with such a non-stationary component (e.g. \cite{Eppley:1977,
Abrahams-Heiderich-etal:1992, Garfinkle-Duncan:2001,
Alcubierre-Allen-etal:2000}).  They are based on certain simplifying
assumptions, and allow for variations of the gravitational wave
perturbation through a freely specifiable function, commonly called
$q$.  Brill's idea has also been generalized to three dimensions and
to black hole spacetimes (e.g., \cite{Brandt-Seidel:1996,
Shibata:1997b, Camarda-Seidel:1999, Baker-Brandt-etal:2000,
Brandt-Camarda-etal:2003, Brown-Lowe:grqc0408089}).  All of these
authors continue to encode the perturbation in a function $q$.  It
appears that, generally, this function is chosen rather ad hoc, its
purpose mainly being to perturb the spacetime in {\em some} way.
While every (nonzero) choice for $q$ leads to a perturbed initial data
set, it is not clear what properties the perturbation has, nor how to
control these properties.  Given that $q$ is often chosen to be
bell-shaped (e.g., a Gaussian), it seems likely that the resulting
perturbation is some vaguely localized lump of energy, rather than,
say, a coherently traveling wave.  Part of the motivation to use Brill
waves was certainly that they lead to fairly simple equations which
are easy to solve numerically.  Since elliptic solvers have matured
considerably over the last years (e.g. \cite{Marronetti-Matzner:2000,
Grandclement-Gourgoulhon-Bonazzola:2001b, Pfeiffer-Kidder-etal:2003,
Tichy-Bruegmann-etal:2003}), computational complexity is no longer a
serious issue, and one is free to look for more general approaches,
with easier to interpret properties of the resulting initial data
sets.

An obvious starting point are linearized gravitational
waves~\cite{Misner-Thorne-Wheeler:1973}, which then are incorporated
into the solution of the initial value
problem~\cite{Abrahams-Evans:1992, Abrahams-Evans:1993,
Shibata-Nakamura:1995, Bonazzola-Gourgoulhon-etal:2004}.  Here, we
continue along this line of thought, and propose a conceptually very
simple method, which combines linear gravitational waves with the
conformal thin sandwich formalism~\cite{York:1999}.  Our basic idea is
to build the linear gravitational wave into the free data for the
conformal thin sandwich equations.  The method allows superposition of
an arbitrary linear gravitational wave onto an arbitrary background
spacetime.  The constructed data sets retain, at least qualitatively,
the properties of the underlying linear gravitational wave.  Thus,
properties of the perturbation to be inserted in the initial data set
can be controlled easily by selecting the appropriate underlying
linear gravitational wave solution.

In Sec.~\ref{sec:Method}, we present the method and discuss its
relationship to previous work~\cite{Abrahams-Evans:1992,
Abrahams-Evans:1993, Shibata-Nakamura:1995,
Bonazzola-Gourgoulhon-etal:2004}.  As an illustration we superpose,
in Sec.~\ref{sec:Results}, quadrupolar gravitational waves on
Minkowski space and on a Schwarzschild black hole.  We close with a
discussion in Sec.~\ref{sec:Discussion}.

\section{Method}
\label{sec:Method}

Employing the usual 3+1 decomposition of Einstein's
equations\cite{Arnowitt-Deser-Misner:1962, York:1979}, the
spacetime metric is written as
\begin{equation}
^{(4)}ds^2=-\N^2 dt^2 + g_{ij}(dx^i+\beta^i dt)(dx^j+\beta^j dt),
\end{equation}
where $g_{ij}$ represents the spatial metric on
$t$=const. hypersurfaces, and $\N$ and $\beta^i$ denote the lapse
function and shift vector, respectively.  The extrinsic curvature,
$K_{ij}$ is defined by $K=-\frac{1}{2}\perp{\cal L}_n ^{(4)}{\bf g}$,
where $^{(4)}{\bf g}$ represents the spacetime metric, $n$ the
future-pointing unit normal to the hypersurface, and $\perp$ the
projection operator into the hypersurface.  Einstein's equations then
split into evolution equations,
\begin{align}
\label{eq:dtgij}
(\partial_t-{\cal L}_\beta )g_{ij}&=-2\N K_{ij},\\
\label{eq:dtKij}
(\partial_t\!-\!{\cal L}_\beta) K_{ij}&=
N\left(R_{ij}\!-\!2K_{ik}K^k{}_j\!+\!K K_{ij}\right)
\!-\!\nabla_{\!i}\nabla_{\!j}\N,
\end{align}
and constraint equations, 
\begin{align}
\label{eq:Ham}
R+K^2-K_{ij}K^{ij}&=0,\\
\label{eq:Mom}
\nabla_j\left(K^{ij}-g^{ij}K\right)&=0.
\end{align}
Here, $\nabla_{\!i}$ is the covariant derivative compatible with
$g_{ij}$, ${\cal L}$represents the Lie-derivative, and $R_{ij}$
denotes the Ricci tensor of $g_{ij}$.  Furthermore, $R$ and $K$ denote
the traces of the Ricci tensor and the extrinsic curvature,
respectively, and we have assumed vacuum.  Initial data for Einstein's
equations consists of $(g_{ij}, K^{ij})$.  The difficulty in
constructing such data lies in the requirement that the Hamiltonian
and momentum constraints, Eqs.~(\ref{eq:Ham}) and~(\ref{eq:Mom}), must
be satisfied.

One widely used formalism for constructing initial data is the
conformal thin sandwich approach~\cite{York:1999, Pfeiffer-York:2003}.
It is based on two neighboring hypersurfaces, their conformal
three-geometries, and the instantaneous time-derivative of the
conformal three-geometry.  One introduces a conformal spatial metric
$\tilde g_{ij}$, related to the physical spatial metric by
\begin{equation}\label{eq:Confg}
g_{ij}=\CF^4\tilde g_{ij},
\end{equation}
where $\psi$ is called the conformal factor.  To construct initial
data, one chooses the conformal metric $\tilde g_{ij}$, its
time-derivative 
\begin{equation}
\tilde u_{ij}=\partial_t\tilde g_{ij},
\end{equation}
as well as the trace of the extrinsic curvature $K$ and the conformal
lapse $\tilde \N=\CF^{-6}\N$.  We note that $\tilde u_{ij}$ must be
traceless, $\tilde u_{ij}\tilde g^{ij}=0$.  Having made these choices,
the Hamiltonian and momentum constraints take the form
\begin{align}\label{eq:Ham3}
\cderiv^2\CF-\frac{1}{8}\CF \cR-\frac{1}{12}\CF^5\trK^2
+\frac{1}{8}\CF^{-7}\cA_{ij}\cA^{ij}=0,\\
\label{eq:Mom3}
\cderiv_j\Big(\frac{1}{2\cN}(\cLong\beta)^{ij}\Big)
-\cderiv_j\Big(\frac{1}{2\cN}\cu^{ij}\Big)-\frac{2}{3}\CF^6\cderiv^i\trK
=0.
\end{align}
Here, $\cderiv_{\!i}$ and $\cR$ are the covariant derivative
compatible with $\tilde g_{ij}$ and the trace of the Ricci tensor of
$\tilde g_{ij}$, respectively, $\cLong$ denotes the
longitudinal operator,
\begin{equation}
(\cLong
\beta)^{ij}=\cderiv^i\beta^j+\cderiv^j\beta^i-\frac{2}{3}\tilde
g^{ij}\cderiv_k\beta^k,
\end{equation}
and $\tilde A^{ij}$ is defined as
\begin{equation}
\tilde A^{ij}=\frac{1}{2\tilde \N}\Big((\cLong\beta)^{ij}-\tilde u^{ij}\Big).
\end{equation}
Equations~(\ref{eq:Ham3}) and~(\ref{eq:Mom3}) are elliptic equations
for the conformal factor $\CF$ and the shift $\beta^i$.  After solving
these equations for $\CF$ and $\beta^i$, the physical initial data 
is given by Eq.~(\ref{eq:Confg}) and by
\begin{equation}
K_{ij}=\CF^{-10}\tilde A^{ij}+\frac{1}{3}g^{ij}K.
\end{equation}

Instead of specifying $\cN$ as part of the free data one can also set
$\partial_tK$.  It is well known that this leads to an elliptic
condition for the lapse-function (e.g. \cite{Smarr-York:1978,
Wilson-Mathews:1989, Pfeiffer-York:2003}):
\begin{align}\nonumber
\cderiv^2(\cN\CF^7)\!-\!(\cN\CF^7)\Big(\frac{1}{8}\cR\!+\!
\frac{5}{12}\CF^4\trK^2
\!+\!\frac{7}{8}\CF^{-8}\cA_{ij}\cA^{ij}\Big)&\\
=-\CF^5\left(\partial_t\trK\!-\!\beta^k\partial_k\trK\right)&.
\label{eq:dtK3}
\end{align}

The second ingredient into the construction of perturbed initial data
is {\em linear} gravitational waves.  In linearized
gravity~\cite{Misner-Thorne-Wheeler:1973}, the spacetime metric is
written as
\begin{equation}\label{eq:gfour-linwave}
\gfour_{\mu\nu}=\eta_{\mu\nu}+A\,h_{\mu\nu},
\end{equation}
where $\eta_{\mu\nu}$ is the Minkowski-metric, $A\ll 1$ a constant,
and $h_{\mu\nu}={\cal O}(1)$ the linear gravitational wave.  (We
separate the amplitude $A$ from $h_{\mu\nu}$ for later convenience.)
In transverse-traceless gauge~\cite{Misner-Thorne-Wheeler:1973}, $h_{\mu\nu}$
is purely spatial, $h_{\mu0}=h_{0\mu}=0$, transverse with respect to
Minkowski space, $\nabla^ih_{ij}=0$, and traceless,
$\eta^{ij}h_{ij}=0$.  To first order in the amplitude $A$, Einstein's
equations reduce to
\begin{equation}\label{eq:box-hij}
\square h_{ij}=0,
\end{equation}
where $\square$ is the Minkowski space d'Alambertian. 
The 3+1 decomposition of the metric (\ref{eq:gfour-linwave}) in
transverse-traceless gauge is
\begin{align}
\label{eq:3+1-linwave-gij}
g_{ij}&=f_{ij}+A\,h_{ij}\\
\beta^i&=0,\\
\N&=1,
\end{align}
where $f_{ij}$ denotes the flat spatial metric.  From the evolution
equation for $\g_{ij}$, Eq.~(\ref{eq:dtgij}), we find the extrinsic
curvature
\begin{equation}\label{eq:3+1-linwave-Kij}
K_{ij}=-\frac{A}{2}\dot h_{ij}.
\end{equation}

The spacetime metric (\ref{eq:gfour-linwave}) satisfies Einstein's
equations to first order in $A$.  Consequently, $(g_{ij}, K^{ij})$
from Eqs.~(\ref{eq:3+1-linwave-gij}) and (\ref{eq:3+1-linwave-Kij})
will satisfy the constraints to linear order in $A$.  Since we intend
to increase $A$ to order unity, this is not sufficient, and we must
solve the constraint equations.  Because the spatial metric,
Eq.~(\ref{eq:3+1-linwave-gij}) and its time-derivative, $A\dot
h_{ij}$, are known, it seems appropriate that this information be
incorporated into the constraint-solve.

In light of the conformal thin sandwich formalism, it seems obvious to
use Eq.~(\ref{eq:3+1-linwave-gij}) as conformal metric, and to base
the time-derivative of the conformal metric on $\dot h_{ij}$:
\begin{align}
\label{eq:cgij-linwave1}
\cg_{ij}&=f_{ij}+A\, h_{ij},\\
\label{eq:cuij-linwave1}
\cu_{ij}&=A\dot h_{ij}-\frac{1}{3}\cg_{ij}\cg^{kl}A\dot h_{kl}.
\end{align}
The second term in (\ref{eq:cuij-linwave1}) ensures that $\tilde
u_{ij}$ is tracefree with respect to $\cg_{ij}$.  Because $h_{ij}$ and
$\dot h_{ij}$ are traceless, Eq.~(\ref{eq:3+1-linwave-Kij}) suggests
the choice
\begin{equation}\label{eq:trK-linwave}
 \trK=0.
\end{equation}
The free data is completed by setting
\begin{equation}\label{eq:trKdot-linwave}
\partial_t\trK=0.
\end{equation}

While the free data
Eqs.~(\ref{eq:cgij-linwave1})--(\ref{eq:trKdot-linwave}) were
motivated by a small perturbation, they can be used equally well for
large amplitudes $A$ (as long as solutions can be found).  Hence, by
increasing $A$, one can obtain nonlinearly perturbed initial data
sets.

In writing down Eqs.~(\ref{eq:trK-linwave})
and~(\ref{eq:trKdot-linwave}), we have neglected terms of order ${\cal
O}(A^2)$ on the right hand sides which arise because $h_{ij}$ is
traceless with respect to the flat metric $f_{ij}$, but not with
respect to the perturbed metric $f_{ij}+A h_{ij}$.  Linearized gravity
cannot determine such higher order terms.  Since nonlinearities of
Einstein's equations arise at the same order, and these nonlinearities
are not accounted for in $h_{ij}$, we see no advantage to including
${\cal O}(A^2)$ terms in Eqs.~(\ref{eq:trK-linwave})
and~(\ref{eq:trKdot-linwave}).  We have also chosen to use
Eq.~(\ref{eq:trKdot-linwave}) as free data and include
Eq.~(\ref{eq:dtK3}) as a fifth elliptic equation.  An
alternative is to set $\cN=1$, and to solve only the four
equations~(\ref{eq:Ham3}) and~(\ref{eq:Mom3}).  Both alternatives are
identical to linear order in $A$.

Equations~(\ref{eq:cgij-linwave1})--(\ref{eq:trKdot-linwave}), which
result in a perturbation of Minkowski space, can be generalized to
curved backgrounds easily by replacing the flat metric by a curved
metric: Let $\g_{ij}^0$ and $\trK^0$ be the 3-metric and mean
curvature of an asymptotically flat, spatial slice through a
stationary spacetime (for example flat space or a Kerr black hole).
Solve the conformal thin sandwich equations (\ref{eq:Ham3}),
(\ref{eq:Mom3}), and~(\ref{eq:dtK3}) with the free data
\begin{align}
\label{eq:cgij-linwave2}
\cg_{ij}&=\g_{ij}^0+A\,h_{ij},\\
\cu_{ij}&=A\,\dot h_{ij}-\frac{1}{3}\cg_{ij}\cg^{kl}A\dot h_{kl},\\
\trK&=\trK^0,\\
\label{eq:dtK-linwave2}
\partial_t\trK&=0.
\end{align}
We consider a few limiting cases
\begin{itemize}
\item For $A=0$ the free data reduce to $\cg_{ij}=g^0_{ij}, K=K^0,
\tilde u_{ij}=\partial_tK=0$.  In this case, the underlying
stationary spacetime is a solution of the conformal thin sandwich
equations. 
\item For $A\ll 1$ and the wave $h_{ij}$ located in the asymptotically
flat region of the hypersurface, linear theory is valid.  The
properties of the perturbation in the initial data set will be
precisely those of the underlying linear wave $h_{ij}$.
\item
For large $A$ we will have a nonlinearly perturbed spacetime,
our primary interest.  Due to the nonlinearity of Einstein's
equations, the properties of such a strongly perturbed spacetime will
differ from the linear wave. However, we expect that the qualitative
properties are unchanged.
\end{itemize}

For constructing perturbed initial data on a curved background, one
can, of course, also use a gravitational wave $h_{ij}$ which
represents a linear wave on the {\em background} $g_{ij}^0$, rather
than on flat space.  In that case, the $A\ll 1$ limit approaches the
underlying linear wave even if the underlying wave is located in the
strong field region.  Since construction of linear waves on curved
backgrounds is more complicated than on flat space, the decision
whether this is necessary for a particular application will depend on
how closely the perturbation must match an exact linear wave in the
limit $A\ll 1$.  Superposition of a flat space linear wave at
intermediate separations from a black hole, say, $10M$ or $20M$,
should result in a gravitational wave which predominantly, although
not exactly, retains the properties of the linear wave, which may be
sufficient for many applications.

Finally, we remark that our approach is related to and generalizes
work by Abrahams \&
Evans~\cite{Abrahams-Evans:1992,Abrahams-Evans:1993}, Shibata \&
Nakamura~\cite{Shibata-Nakamura:1995} and Bonazzola {\em et
al.}~\cite{Bonazzola-Gourgoulhon-etal:2004}.  Abrahams \&
Evans~\cite{Abrahams-Evans:1992,Abrahams-Evans:1993} assume
axisymmety, and set a certain component of the extrinsic curvature
(namely $K^r_{\;\theta}$ in spherical coordinates) equal to the value
appropriate for the linear wave.  Then they solve the momentum
constraints for the remaining components of $\K_{ij}$.  This procedure
singles out a preferred coordinate system, while our method is
covariant with respect to spatial coordinate transformations.

Shibata \& Nakamura~\cite{Shibata-Nakamura:1995} use the extrinsic
curvature decomposition~\cite{York:1979,Pfeiffer-York:2003} to
construct initial data, rather than the conformal thin sandwich
equations.  By choosing a maximal slice, $\trK=0$, Hamiltonian and
momentum equation decouple, and the momentum constraint is solved by
the analytical (transverse tracefree) extrinsic curvature determined
from the underlying linearized wave.  This is a very elegant approach,
since only the Hamiltonian constraint remains to be solved for the
conformal factor; however, decoupling of the Hamiltonian and momentum
constraint happens only for slices with constant $\trK$.  In order to
construct perturbations of black hole spacetimes with non-constant
$\trK$, the coupled elliptic equations within the extrinsic curvature
decomposition have to be solved.  This extension is conceptually
straightforward and would lead to a method parallel to the one
presented here, but using the extrinsic curvature decomposition rather
than the conformal thin sandwich formalism.

Bonazzola {\em et al.}~\cite{Bonazzola-Gourgoulhon-etal:2004},
finally, uses the conformal thin sandwich formalism to superpose a
time-symmetric linearized wave ($\partial_t\cg_{ij}=0$, i.e. a
superposition of an incoming and an outgoing wave) on flat space.
This work employs Dirac-gauge and relies heavily on spherical
coordinate systems in the reduction of the remaining degrees of
freedom to two scalar functions.  Our method, in contrast, can be used
with any spatial coordinates, which is of particular importance for
superposition of gravitational waves on black hole backgrounds, which
may not be available in Dirac-gauge.  Furthermore, while we use
Teukolsky waves below as an example, our method can also applied to
different linearized waves, for example a spherical wave superposed
with a plane wave, which would be more difficult to implement in the
approach taken in Ref.~\cite{Bonazzola-Gourgoulhon-etal:2004}.

\section{Numerical results}
\label{sec:Results}

\subsection{Quadrupole waves}

We illustrate the general method introduced in Sec.~\ref{sec:Method}
with linearized quadrupole waves as given by
Teukolsky~\cite{Teukolsky:1982}.  This reference explicitly presents
even parity waves, which are superpositions of $l=0,2,$ and $4$ modes,
as well as odd parity waves, which are constructed as superpositions
of $l=1$ and $3$.  For each parity, there are five independent modes,
corresponding to azimuthal quantum number $m=\pm 2, \pm 1, 0$.  The
even parity outgoing wave has a spacetime line-element
\begin{equation}\label{eq:TeukolskyWave}
\begin{aligned}
^{(4)}ds^2=-dt^2&+(1+Af_{rr})dr^2+(2Bf_{r\theta})rdrd\theta\\
&+(2Bf_{r\theta})r\sin\theta drd\phi\\
&+\left(1+Cf_{\theta\theta}^{(1)}+Af_{\theta\theta}^{(2)}+\right)r^2d\theta^2\\
&+\left[2(A-2C)f_{\theta\phi}\right]r^2\sin\theta d\theta d\phi\\
&+\left(1+Cf_{\phi\phi}^{(1)}
+Af_{\phi\phi}^{(2)}\right)r^2\sin^2\theta d\phi^2.
\end{aligned}\end{equation}
with radial dependence given by
\begin{align}
\label{eq:A}
A&
=3\left[\frac{F^{(2)}}{r^3}\!+\!\frac{3F^{(1)}}{r^4}\!+\!\frac{3F}{r^5}\right],
\\
B&=-\left[\frac{F^{(3)}}{r^2}\!+\!\frac{3F^{(2)}}{r^3}
\!+\!\frac{6F^{(1)}}{r^4}+\frac{6F}{r^5}\right],\\
\label{eq:C}
C&=\frac{1}{4}\left[\frac{F^{(4)}}{r}\!+\!\frac{2F^{(3)}}{r^2}
\!+\!\frac{9F^{(2)}}{r^3}+\frac{21F^{(1)}}{r^4}\!+\!\frac{21F}{r^5}\right],
\end{align}
where
\begin{equation}
\label{eq:F-deriv}
F^{(n)}q\equiv \left[\frac{d^nF(x)}{dx^n}\right]_{x=t-r}.
\end{equation}
$F(x)=F(t-r)$ describes the shape of the wave.  The functions $f_{rr},
\ldots, f^{(2)}_{\phi\phi}$ depend only on angles $(\theta,\phi)$;
they are given explicitly in Ref.~\cite{Teukolsky:1982} for each
azimuthal quantum number $M$.  Ingoing quadrupole waves are obtained
by replacing $F(t-r)$ with a function of $t+r$, and reversing the
signs in front of odd derivatives of $F$ in Eq.~(\ref{eq:F-deriv}).
Reference~\cite{Teukolsky:1982} gives also the metric for odd parity
waves.  From Eq.~(\ref{eq:TeukolskyWave}), one can easily extract
$h_{ij}$ and $\dot h_{ij}$.

\subsection{Flat space with ingoing pulse}

We consider a perturbation of flat space,
$g_{ij}^0=f_{ij}$, $\trK^0=0$.  We choose the even parity, $m=0$
ingoing mode.  The shape of the pulse is taken as a Gaussian
\begin{equation}\label{eq:linwave-Gaussian}
F(x)=e^{-(x-x_0)^2/w^2}
\end{equation}
of width $w=1$ and with an initial radius of $x_0=20$.

\begin{figure}
\centerline{\includegraphics[scale=0.35]{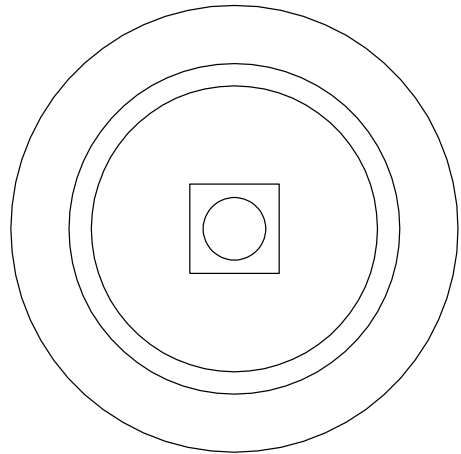}} 
\caption{\label{fig:GW-domain}Domain decomposition in $R^3$.  A cube
covers the central region which is not covered by the spherical
shells.}
\end{figure}

Equations~(\ref{eq:Ham3}),~(\ref{eq:Mom3}), and~(\ref{eq:dtK3}) are
solved with the pseudospectral elliptic solver described in
\cite{Pfeiffer-Kidder-etal:2003}.  The domain decomposition used in
the elliptic solver is shown in Figure~\ref{fig:GW-domain}.  We use
three spherical shells with boundaries at radii $r=1.5, 16, 24,$ and
$10^9$, so that the middle shell is centered on the gravitational wave.
The inner shell does not extend to the origin, since the regularity
conditions at the origin of a sphere are not implemented in the code.
Instead, we place a cube on the origin which overlaps the innermost
spherical shell.  The solutions of the constraint equations turn out
to be very smooth except for high frequency radial variations at the
location of the gravitational wave pulse,
cf. Eq.~(\ref{eq:linwave-Gaussian}).  Therefore, the accuracy is
completely determined by the number $N_r$ of radial basis functions in
the middle spherical shell, which is chosen significantly larger than
the number of basis functions in the other subdomains.  At the highest
resolution, there are $N_r=84$ radial basis functions in the middle
shell, but only $28$ in the other two shells.  Furthermore, the
angular resolution of all shells is $L=15$ and the cube has $18$
basis functions in each dimension.

\begin{figure}
\centerline{\includegraphics[scale=0.36]{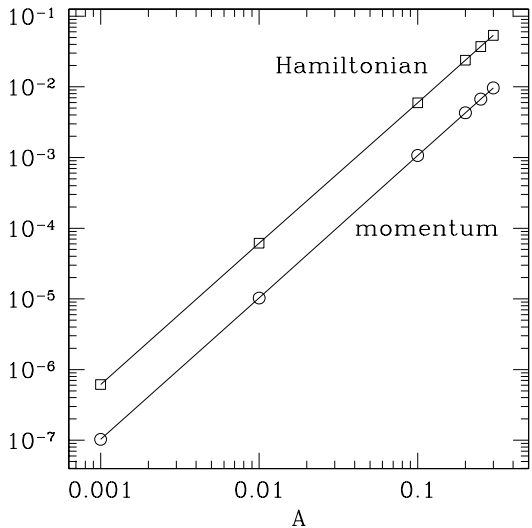}}
\caption{\label{fig:GW-violation}Constraint violation of linear
gravitational wave in flat background prior to solving the
constraints.}
\end{figure}

Figure~\ref{fig:GW-violation} presents the residuals of Hamiltonian
and momentum constraints, Eqs.~(\ref{eq:Ham}) and~(\ref{eq:Mom}) for
the linear gravitational wave {\em without} solving the constraints,
i.e.  upon direct substitution of Eqs.~(\ref{eq:3+1-linwave-gij}) and
(\ref{eq:3+1-linwave-Kij}) into the constraint equations.  As
expected, the residual is ${\cal O}(A^2)$, confirming that the
quadrupole wave is indeed a solution of linearized gravity.

\begin{figure}
\centerline{\includegraphics[scale=0.44]{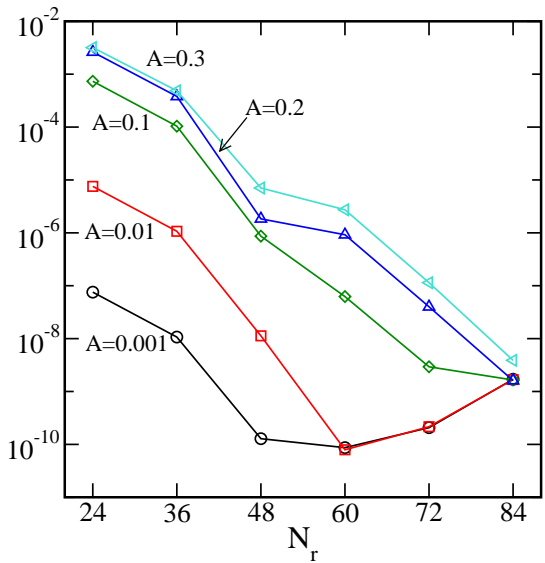}}
\caption{\label{fig:GW-flatspace-conv1} Convergence of the
elliptic solver for different amplitudes $A$.  Plotted is the residual
in the Hamiltonian constraint (root mean square) vs. the number of
radial basis functions in the middle spherical shell.  }
\end{figure}

\begin{figure}
\centerline{\includegraphics[scale=0.44]{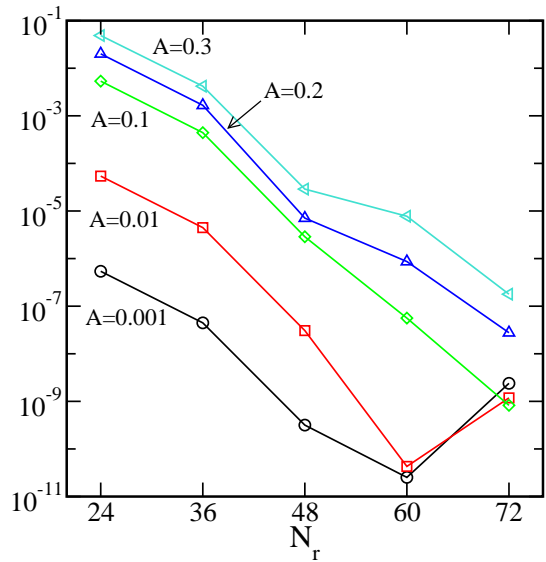}}
\caption{\label{fig:GW-flatspace-conv2}Convergence of the
elliptic solver for different amplitudes $A$.  Plotted is the
difference of the ADM energy to the next higher resolution solution
vs. the number of radial basis functions in the middle spherical
shell.  }
\end{figure}

\begin{figure}
\centerline{\includegraphics[scale=0.36]{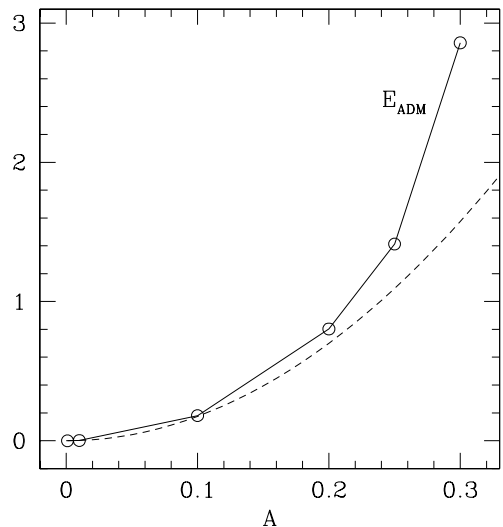}}
\caption{\label{fig:GW-flatspace-E}ADM energy of an ingoing Gaussian
pulse in flat space.  The dashed line indicates the low-amplitude
quadratic behavior.}
\end{figure}

We now solve the conformal thin sandwich equations with the free
data~(\ref{eq:cgij-linwave2})--(\ref{eq:dtK-linwave2}) for different
$A$, and compute the ADM energy for each solution,
\begin{equation}
E_\mathrm{ADM}=\frac{1}{16\pi}\int_{\infty}
  \left(g_{ij,j}-g_{jj,i}\right)\,d^2S_i.
\end{equation}
Figures~\ref{fig:GW-flatspace-conv1} and~\ref{fig:GW-flatspace-conv2}
plot the residual of the Hamiltonian constraint and the error of the
ADM energy versus $N_r$.  Exponential convergence is apparent, until
roundoff limit is encountered around $\sim 10^{-10}$.  The seemingly
large value of this number (when compared to the usual double
precision floating point accuracy of $\sim 10^{-17}$) is a consequence
of the many numerically computed derivatives that enter the
calculation: To compute $\tilde R$ in Eq.~(\ref{eq:Ham}), second
numerical derivatives of the conformal metric
Eq.~(\ref{eq:3+1-linwave-gij}) are taken.  After solution of the
elliptic equations, the physical metric $\g_{ij}$ is assembled, and
the Hamiltonian constraint Eq.~(\ref{eq:Ham}) is evaluated using
second numerical derivatives of $g_{ij}$.  Each of these numerical
differentiations increases the roundoff error by a factor of order the
number of basis functions.  The increase in roundoff error with the
number of basis functions can be clearly seen in Fig. 3

We now turn our attention from the convergence properties to the
actual solutions of the constraint equations.  For small $A$, we find
that $\CF-1$ is proportional to $A^2$ everywhere.  This is expected,
because $\CF-1$ corrects the conformal metric to satisfy the
Hamiltonian constraint.  As the constraint violation is proportional
to $A^2$, so is this correction.  Fig.~\ref{fig:GW-flatspace-E} proves
that one can clearly achieve initial data sets with a significant
energy content.  At low amplitudes, $E_\mathrm{ADM}$ is proportional
to $A^2$, as one expects given that $\CF-1$ is proportional to $A^2$.
At high amplitudes, however, $E_\mathrm{ADM}$ grows faster than $A^2$,
indicating that the non-linear regime with self-interaction is
reached.  For $A>0.3$, the elliptic solver fails to converge.

\begin{figure}
\centerline{\includegraphics[scale=0.40]{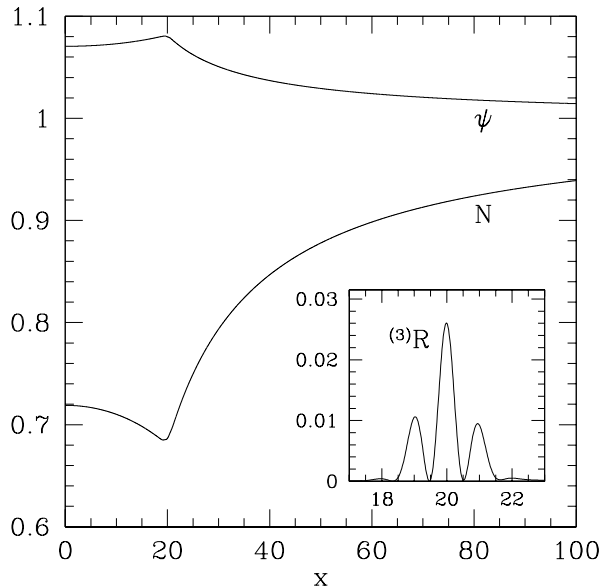}}
\caption{\label{fig:GW-flatspace-cuts}Cuts through the equatorial
plane of the $A\!=\!0.3$ data set of
Fig.~\ref{fig:GW-flatspace-E}. The large plot shows lapse and
conformal factor, the inset shows the scalar curvature of the
3-metric.}
\end{figure}

We now discuss the data set with amplitude, $A=0.3$ in more detail.
Its ADM energy is $E_\mathrm{ADM}=2.858$.
Figure~\ref{fig:GW-flatspace-cuts} presents cuts through the conformal
factor $\CF$, lapse $\N$ and the scalar curvature of the physical
3-metric, $^{(3)}\R$.  Conformal factor and lapse deviate
significantly from unity confirming that the solution is indeed deep
in the nonlinear regime.  The scalar curvature is virtually zero
everywhere except within a spherical shell with $18\lesssim r\lesssim
20$.  Although the linearized wave is based on a Gaussian profile
Eq.~(\ref{eq:linwave-Gaussian}), even finer features are introduced
into the linearized wave (and into the scalar curvature plotted in
Fig.~\ref{fig:GW-flatspace-cuts}) because of the derivatives in
Eqs.~(\ref{eq:A})--(\ref{eq:C}).  Resolution of these fine features
necessitates the high radial resolution of $N_r$.

The gravitational wave is concentrated in a spherical shell of width
$w=1$.  The underlying {\em linear} wave is purely ingoing, so it
seems reasonable that the gravitational perturbation in the physical,
{\em nonlinear} spacetime is also predominantly ingoing.  Neglecting
dispersion, the wave will concentrate in a sphere centered at the
origin with radius $r\sim w$.  Black holes usually form for systems
with mass to size ratio of order unity; here, $E_\mathrm{ADM}/w\approx 2.8$,
so that black hole formation appears very likely once the pulse is
concentrated at the origin.

These data sets could be used to examine critical collapse to a black
hole, repeating Abrahams \& Evans~\cite{Abrahams-Evans:1993} and
extending it to genuinely three-dimensional collapse by choosing
$m\neq 0$ in the underlying quadrupole wave.  These datasets also
provide a testbed for evolution codes in situations where the {\em
topology} of the horizons changes.

\subsection{Black hole with gravitational wave}
\label{sec:GW:BH}

As a second example of the flexibility of our method, we superpose
a gravitational wave on a black hole background.  The background
spatial metric and trace of the extrinsic curvature are set to a
Schwarzschild black hole in Eddington-Finkelstein coordinates,
\begin{align}
\label{eq:gij-EF}   \g^{0}_{ij}&=\delta_{ij}+\frac{2M}{r}n^in^j, \\
\label{eq:trK-EF}
\trK_{0}&=\frac{2M}{r^2}\left(1+\frac{2M}{r}\right)^{-3/2}
      \left(1+\frac{3M}{r}\right).
\end{align}
where $n^i=x^i/r$, and $r^2=\delta_{ij}x^ix^j$.  

We choose an odd, ingoing $m=0$ quadrupole wave with Gaussian shape,
Eq.~(\ref{eq:linwave-Gaussian}) at location $x_0=15$ and width $w=1$.
The metric is singular at the origin, therefore we excise at an inner
radius of $1.5M$ (which is inside the horizon).  At this inner
boundary, we impose simple Dirichlet boundary conditions appropriate
for the unperturbed black hole: $\CF=1$, and $\N=N_0$ and
$\beta^i=\beta^i_0$, with lapse and shift for Eddington-Finkelstein
given by
\begin{align}
\label{eq:lapse-EF} \N_{0}&=\left(1+\frac{2M}{r}\right)^{-1/2},\\
\label{eq:shift-EF}
\beta^i_{0}&=\left(1+\frac{2M}{r}\right)^{-1}\frac{2M}{r}n^i.
\end{align}

\begin{figure}
\centerline{\includegraphics[scale=0.44]{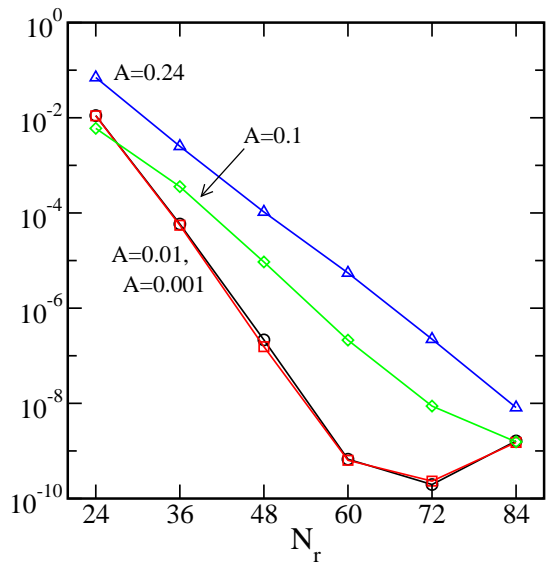}}
\caption{\label{fig:Convergence-BH}Black hole with superposed
gravitational wave: Residual of the Hamiltonian constraint vs. radial
number of basis functions in middle spherical shell.}
\end{figure}

Perturbed initial data sets are constructed for various values of $A$,
and Fig.~\ref{fig:Convergence-BH} demonstrates convergence of the
solutions.  In this case, the resolution is determined by two factors,
namely how well the gravitational wave is resolved in the middle
shell, and how well the inner shell resolves the background solution
$(g^0_{ij}, K^0_{ij})$.  For perturbations of Minkowski space, the
latter was trivial (any expansion resolves the constant Minkowski
background), however, here it is the limiting factor for small $A$ or
low resolutions, whereas for large $A$ and high resolutions, both
effects are about equally important.  At the highest resolution, the
number of radial basis functions in each shell is (from inner to
outer) $48, 84$ and $28$, and the angular resolution is unchanged from
before, $L=15$.  In each resulting initial data set, the apparent
horizon is located with the apparent horizon finder implemented and
tested in \cite{Baumgarte-Cook-etal:1996,
Pfeiffer-Teukolsky-Cook:2000, Pfeiffer-Cook-Teukolsky:2002}, and the
apparent horizon mass is computed from the area of the apparent
horizon via
\begin{equation}
M_{\mathrm{AH}}=\sqrt{\frac{A_\mathrm{AH}}{16\pi}}.
\end{equation}

\begin{figure}
\centerline{\includegraphics[scale=0.40]{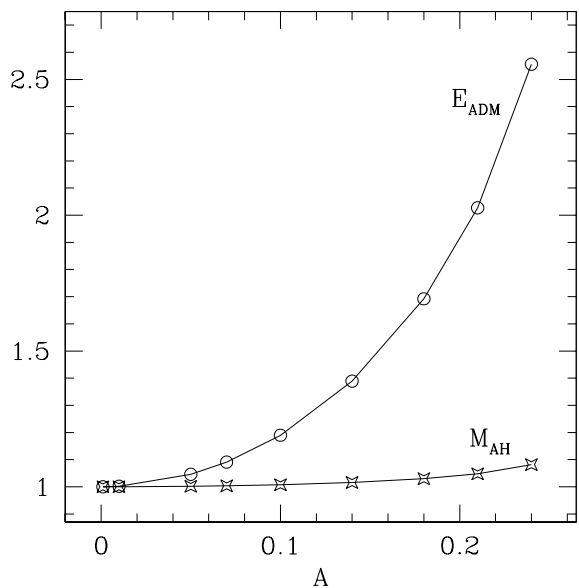}}
\caption{\label{fig:GW-blackhole-E}Black hole with
superposed gravitational wave.}
\end{figure}

Figure~\ref{fig:GW-blackhole-E} presents the ADM energy and the
apparent horizon mass of the central black hole as a function of the
amplitude of the gravitational wave.  The apparent horizon mass is
fairly independent of $A$ indicating that the horizon of the central
black hole is only slightly perturbed by the gravitational wave.
However, the ADM energy, which measures the total energy in the
hypersurface, depends strongly on $A$; for large $A$,
\begin{equation}
\frac{M_\mathrm{ADM}}{M_\mathrm{AH}}\gtrsim 2.5,
\end{equation}
indicating that a significant amount of gravitational energy resides
outside the black hole.

\begin{figure}
\centerline{\includegraphics[scale=0.4]{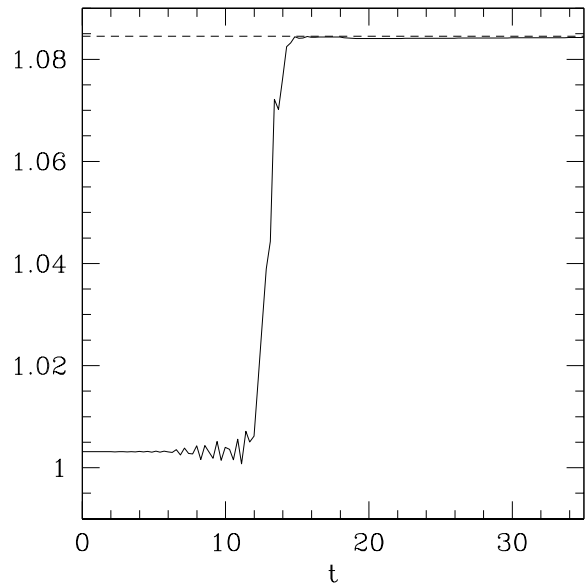}}
\caption{\label{fig:GW-AHmass}Apparent horizon mass during an
evolution of a perturbed black hole spacetime.  The dashed line
indicates $E_\mathrm{ADM}$ as computed from the initial data set.}
\end{figure}

To support our assertion that the superposed initial data set retains
the features of the underlying gravitational wave, we present a
preliminary evolution of a black hole with superposed ingoing
gravitational wave~\cite{Kidder-Shoemaker:2003}.  The initial data for
the evolution is identical to the data sets used in
Figure~\ref{fig:GW-blackhole-E} with the one exception that the
gravitational wave has even parity.  Figure~\ref{fig:GW-AHmass} shows
the apparent horizon mass as a function of evolution time.  All
quantities are scaled such that the unperturbed Schwarzschild black
hole has unit mass.  For $t\lesssim 10$, $M_\mathrm{AH}$ is constant,
its value being that from the initial data set.  Between $12\lesssim
t\lesssim 15$, $M_\mathrm{AH}$ increases rapidly to an asymptotic
value of $M^f_\mathrm{AH}\approx 1.084$.  The ADM energy of the
initial data set was $E_\mathrm{ADM}=1.0845$.  Apparently, the ingoing
gravitational wave outside the black hole falls into it, increasing
the area of the apparent horizon.  The final apparent horizon mass is
very close to the ADM energy, and the growth of $M_\mathrm{AH}$
happens during a time-interval comparable to the width of the initial
pulse.  Thus it appears that a large fraction of the wave is
coherently ingoing and falls into the black hole.

\section{Discussion}
\label{sec:Discussion}

We propose a conceptually very clear method to construct spacetimes
containing gravitational radiation which combines the conformal thin
sandwich formalism with linear gravitational waves.  For small
amplitudes, the gravitational perturbation in the resulting initial
data sets retains the characteristic features of the underlying linear
wave, allowing for easy control of the properties of the gravitational
wave perturbation.  For strong amplitudes, nonlinearities of Einstein's
equations are important, but we expect that the solutions still retain
qualitatively the properties of the underlying linear wave.

To illustrate the method, we superpose quadrupolar gravitational waves
onto Minkowski space, and onto a Schwarzschild black hole.  In both
cases, initial data  with a large amount of gravitational energy in the
perturbation can be constructed.

Numerically, these initial data sets provide test-beds of evolution
codes in situations away from stationarity.  The mass of a central
black hole changes --it may even double-- when a large gravitational
wave falls into it; can current gauge conditions handle this
situation?  If a gravitational wave collapses to a black hole,
horizons appear, and evolution codes using black hole excision must
accommodate this change.  Furthermore, spacetimes with outgoing
gravitational wave perturbations are ideal test-beds for gravitational
wave extraction algorithms, or constraint preserving boundary
conditions~\cite{Kidder-Lindblom-etal:2004}.

Physically, ingoing gravitational wave pulses in Minkowski space, like
the ones presented in Figs.~\ref{fig:GW-flatspace-E}
and~\ref{fig:GW-flatspace-cuts}, could be used to examine critical
collapse, including the genuinely three-dimensional regime with $m\neq
0$.  The black hole initial data sets with ingoing gravitational wave
pulses (cf. Fig.~\ref{fig:GW-blackhole-E}) would be useful to examine
scattering of the gravitational wave at the black
hole~\cite{Allen-Camarda-Seidel:grqc9806014, Papadopoulos:2002,
Zlochower-Gomez-etal:2003, Shoemaker-etal:2004}: What fraction of the
gravitational wave is scattered and reaches infinity?  Which multipole
moments are excited in this process?  This example can also be
generalized to spinning black holes, off-centered gravitational waves,
or gravitational waves with $m\neq 0$.  Interesting questions in these
scenarios would include, whether one can impart linear or angular
momentum on the black hole.

\acknowledgments It is a pleasure to acknowledge helpful discussions
with Lee Lindblom, Saul Teukolsky and Jimmy York.  This work was
supported in part by NSF grants PHY-9900672 at Cornell, PHY-0099568
and PHY-0244906 at Caltech and PHY-0354821 at Penn State.  DS
acknowledges the support of the Center for Gravitational Wave Physics
funded by NSF PHY-0114375.



\end{document}